\documentclass[onecolumn,11pt]{revtex4}  
 
\usepackage[portuguese]{babel}
\usepackage{amsmath}
\usepackage{graphicx}
\usepackage{ulem} 
\usepackage{xcolor}
 \usepackage{hyperref}

\begin{document}

\title{Eventos   científicos de Física no Brasil: 
participação feminina}

\author{Celia Anteneodo$^1$, Carolina Brito$^2$, D\'ebora P. Menezes$^3$}
\address{$^1$Departamento de Física, PUC-Rio, Rio de Janeiro, RJ}

\address{$^2$Instituto de Física da UFRGS, 
 Porto Alegre, RS}

\address{$^3$Departamento de Física - CFM - Universidade Federal de Santa Catarina, Florianópolis, SC}

\begin{abstract}

	We have carried out a partial survey on the female participation in events organized by the Brazilian Physical Society (SBF, in Portuguese) from 2005 to 2021. Our analysis shows that the participation of women has increased over the years, reaching in some areas, the same percentage as the one observed in the SBF community (always lower that 25\%). However, the participation as members of organizing committees and as keynote speakers is always lower. Some propositions to change this inequality situation are listed. \\ 
Realizamos um levantamento parcial sobre a participação de pessoas de  gênero feminino em eventos de Física organizados pela Sociedade Brasileira de Física de 2005 a 2021. 
A nossa análise mostra que a participação feminina nesses eventos aumentou ao longo dos anos, chegando a ter em algumas áreas uma   proporção similar à observada na comunidade da SBF toda (sempre inferior a 25\%). No entanto,   em comitês de eventos, e mais ainda como palestrantes, essa participação é  menor que no conjunto total de participantes. 
Algumas propostas para mudar esse quadro de  desigualdade são elencadas.
\end{abstract}

\maketitle

\section{Introdução}

Grande número de trabalhos, como por exemplo~\cite{Gutterl_2014, McKinsey, Philips_2014,wooley15}, revelam quanto a diversidade é importante na ciência, tanto por  torná-la melhor quanto mais inclusiva. 
 A  diversidade permite trazer novos e distintos pontos de vista  e elaborar perguntas de pesquisa inovadoras, assim enriquecendo o desenvolvimento científico. 
Portanto, em eventos científicos, onde muitas novas ideias são inspiradas, é fundamental que haja diversidade de raça, etnia, religião, sexo, gênero, idade, origem geográfica, etc. 
Além deste motivo fundamental, existem outras 
razões para se preocupar com a questão. 
 Por um lado, uma visível falta de diversidade passa aos potenciais participantes a mensagem de que certos grupos não pertencem à comunidade, assim naturalizando essa exclusão e desestimulando a  participação desses setores, o que realimenta o problema.  
Por outro lado, dado que ministrar boas palestras aumenta o reconhecimento acadêmico e é importante para o processo de promoção na carreira,  quando um grupo  deixa de ser convidado, ou de participar, terá menor visibilidade e  seu desenvolvimento profissional resultará penalizado.

Dada a importância do tema, nosso propósito é estudar o nível de  diversidade em eventos científicos promovidos pela Sociedade Brasileira de Física (SBF).
A SBF foi criada em 1966, durante a XVIII Reunião Anual da Sociedade Brasileira para o Progresso da Ciência (SBPC) em Blumenau, Santa Catarina. Desde então, uma das suas principais atividades tem sido a organização de eventos nacionais temáticos.
Ao longo das décadas, houve algumas mudanças no escopo e na duração desses encontros, mas eles  representam a principal ação para propiciar o desenvolvimento das respectivas áreas. 

\begin{table}[b!]
    \centering
    \begin{footnotesize}
    \begin{tabular}{l|c|c|c|c}
     Nome do evento & Sigla & Média anual  de  &\%  & Período \\[-3mm]
            &       & participantes    &  feminino  &  \\
    \hline   \hline
   \begin{tabular}{l} Encontro Nacional de Física 
   \\[-3mm]  da Matéria Condensada  -
   \\[-3mm]  Encontro de Outono da SBF
   \end{tabular} 
   & \begin{tabular}{c}  ENFMC\\EOSBF \end{tabular} & 958 & 24 & 2005-2021\\    \hline
   \begin{tabular}{l} Simpósio Nacional  de  
   \\[-3mm] Ensino de Física\end{tabular} 
   &SNEF & 1150 & 38 & 2005-2021\\    \hline
   \begin{tabular}{l} Encontro Nacional   de 
   \\[-3mm] Física de Partículas e Campos \end{tabular}    
    &ENFPC& 232 & 18  & 2005-2021\\ \hline

    \begin{tabular}{l} Reunião de trabalho sobre
    \\[-3mm] Física Nuclear no Brasil\end{tabular}    
     & cRTFNBc & 129 & 23  & 2005-2021 \\ \hline
     %
    \begin{tabular}{l} Encontro de Física do
    \\[-3mm]  Norte e Nordeste \end{tabular}    
   &EFNNE & 560 & 25  & 2005-2021 
  \\ \hline
           \begin{tabular}{l} Encontro de Física do
    \\[-3mm]  Norte e Nordeste -Ensino\end{tabular}    
&  EFNNE-Ensino & 447 & 35  & 2010-2015 \\ \hline   
    \begin{tabular}{l} Encontro de Pesquisa em 
    \\[-3mm] Ensino de Física \end{tabular}    
    & EPEF & 267 & 41  & 2016-2020\\ \hline
    \begin{tabular}{l}Escola de Verão
    \\[-3mm] Jorge André Swieca 
    \\[-3mm]
    Física Nuclear Teórica
    \end{tabular}    
    & EVJAS-FNT & 58 & 13  & 2015-2021\\ \hline
    \begin{tabular}{l} Escola de Verão
    \\[-3mm] Jorge André Swieca 
    \\[-3mm]
    Física Nuclear Experimental
    \end{tabular}    
    & EVJAS-FNE & 24 & 29  & 2016-2020\\ \hline
    \begin{tabular}{l} Escola de Verão
    \\[-3mm] Jorge André Swieca 
    \\[-3mm]
    Partículas e Campos
    \end{tabular}    
    & EVJAS-PC & 91 & 19  & 2015-2021\\ \hline
    \begin{tabular}{l}Escola Brasileira de
    \\[-3mm]  Estrutura Eletrônica 
    \end{tabular}
    & EBEE & 120 & 36 & 2016-2021 \\
    \hline\\
      \end{tabular}
        \end{footnotesize}
\caption{Eventos do calendário oficial de SBF:   número médio de participantes por ano em cada evento, e percentual médio da participação feminina em cada evento, dentro do período indicado na última coluna.}
\label{tab:eventos}
\end{table}

O foco deste trabalho é analisar a participação   feminina nos eventos científicos da comunidade de física no Brasil,  dentro do período 2003-2022. A análise é feita  a partir dos dados que constam nos registros da SBF, obtidos a partir dos formulários de inscrição preenchidos para participar em eventos da SBF. Nestes formulários, aparece atualmente como uma das perguntas e suas opções de resposta: ``Gênero: Masculino, Feminino, Outro''. Somente em duas oportunidades foi registrada a opção ``Outro'', e esta   não constava  antes de 2021. Além disso era perguntado ``Sexo'' em vez de ``Gênero''.  Portanto, a nossa análise ficará restrita às categorias ``Masculina'' e ``Feminina'', sem outra distinção por sexo ou gênero, 
não tendo acesso a dados sobre outros marcadores de diversidade de gênero~\cite{PRPER}.

 Como proporção de referência, usaremos o percentual de sócias, em cada ano do período estudado. Veremos como a participação feminina nas diversas atividades dos eventos (participante em quaisquer das atividades, membro de comitês e palestrante) tem mudado  ao longo dos anos, em diversas áreas da Física. 
 Na última seção, discutimos ações que podem ser realizadas, em nível individual, coletivo  e institucional,  para mudar este cenário de desigualdade.

 \section{Eventos científicos da SBF}
  
Uma lista completa dos diversos eventos científicos que compõem o calendário oficial da SBF pode ser encontrada na página web da SBF~\cite{eventosSBF}.  Na Tabela~\ref{tab:eventos}, mostramos a lista destes eventos, que inclui  encontros temáticos, um único encontro regional e escolas temáticas destinadas principalmente a alunos de pós-graduação.
Para cada evento, a tabela apresenta
 o número  médio anual, levando em conta todos os participantes ao longo do período disponível em cada caso,  e o percentual  de  participantes do gênero feminino no mesmo período. Em alguns eventos, o número de participantes sofreu variações consideráveis durante o período estudado, e portanto, as médias devem ser vistas apenas como indicativo do tamanho de cada comunidade que participa dos eventos. \\[1cm]

\begin{figure}[b!]
\centering
\includegraphics[width=0.7\textwidth]{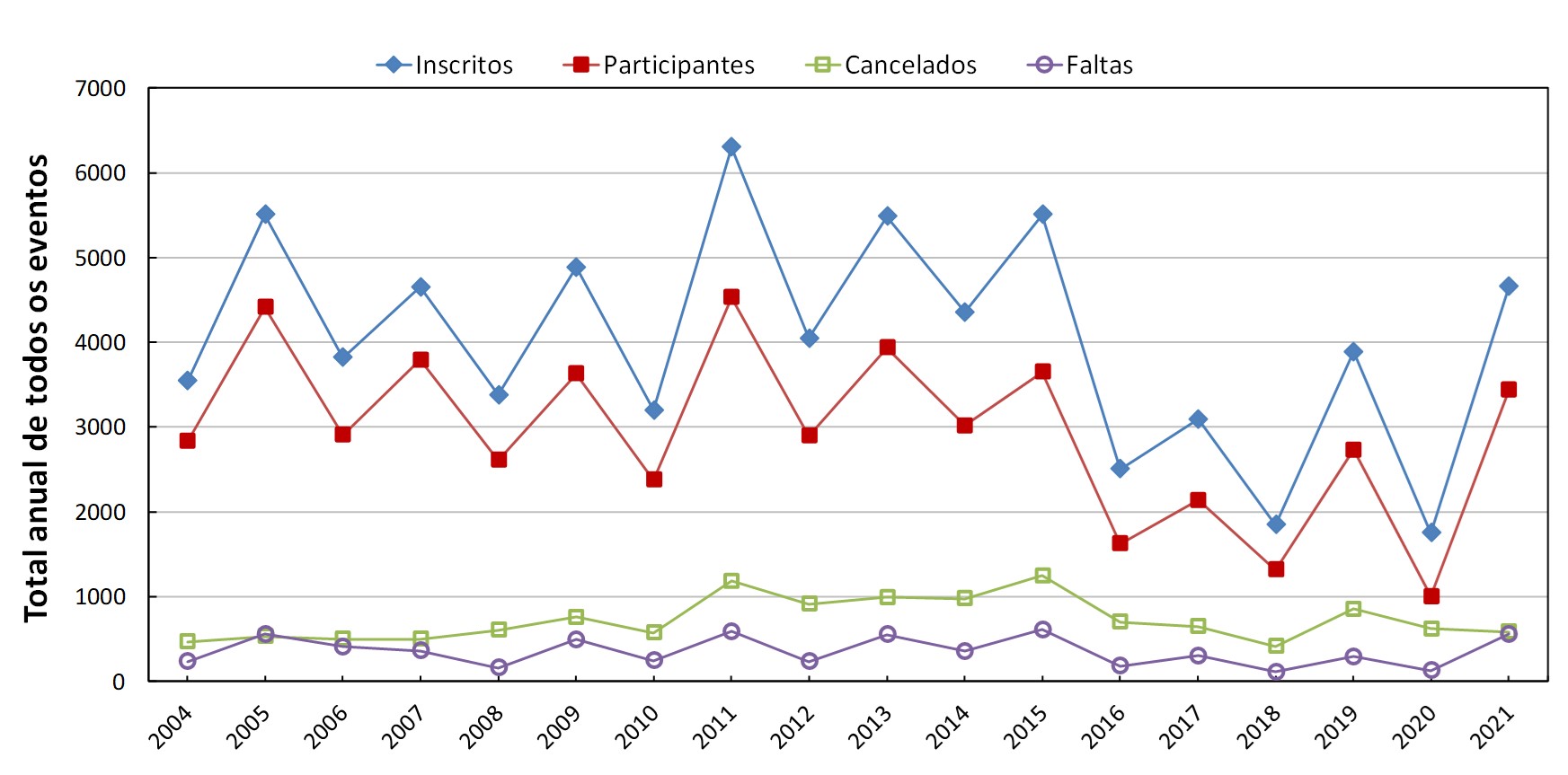}
\caption{ Total anual de participantes, acumulado sobre todos os eventos da SBF a cada ano, no período 2004-2020. Totais de inscritos, participantes efetivos, cancelamentos e faltas são exibidos. }
\label{fig:participantesALL}
\end{figure}

Na Fig.~\ref{fig:participantesALL}, mostramos a evolução temporal do número total de participantes nos diversos eventos da SBF, agrupando os dados por ano.  As oscilações se devem ao fato de alguns eventos, como
o SNEF e as Escolas de Verão Jorge André Swieca,  serem bienais, e outros, como os EFNNEs, não terem sido realizados todo ano. 

O número efetivo de participantes corresponde aos inicialmente inscritos menos  cancelamentos e faltas, números também mostrados Fig.~\ref{fig:participantesALL}.   Apesar das grandes oscilações, vê-se uma clara tendência de diminuição na participação nestes eventos nacionais nos últimos anos.
Esporadicamente (2011 e 2016), foi realizado o Encontro Nacional de Física (familiarmente, Encontrão) que reúne todas as áreas, entretanto, cada evento de área dentro do Encontrão conserva seus registros próprios, permitindo uma análise separada de cada um deles.

\begin{figure}[b!]
\centering
\includegraphics[width=0.6\textwidth]{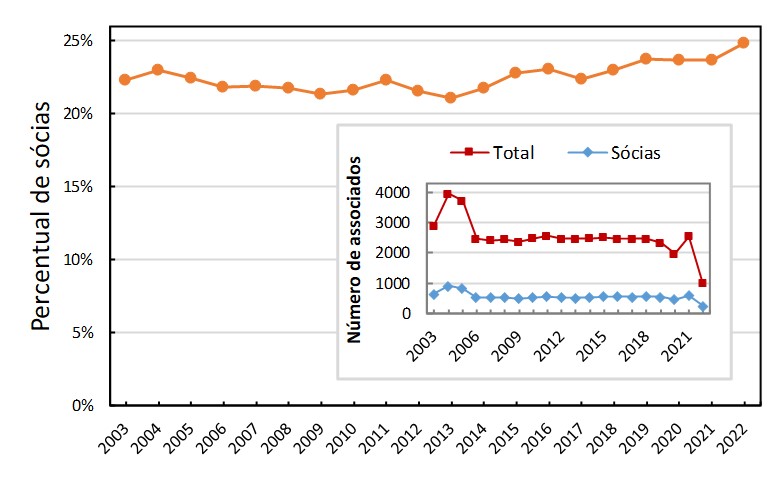}
\caption{Percentual de associadas à SBF, no período 2003-2022.  No gráfico inserido é mostrado o número total de sócios adimplentes de ambos os sexos, e o número de sócias adimplentes. 
}
\label{fig:socias}
\end{figure}

Como referência,  para estimar a composição por gênero da comunidade da SBF, apresentamos na Fig.~\ref{fig:socias} o percentual de sócias ao longo do período estudado. 
No gráfico inserido mostramos também  o número total de  associados à SBF (adimplentes de ambos os sexos considerados)  e  o número total de sócias adimplentes. Observamos uma tendência de redução do número total de sócios nos últimos anos, exceto picos circunstanciais (geralmente vinculados à realização de grandes encontros de física), 
 enquanto no subconjunto de sócias a redução foi menor, de tal forma que a proporção feminina apresenta um leve aumento nos últimos anos.  
Entretanto, este aumento pode estar relacionado a uma mudança da composição da comunidade de sócios, com a incorporação de colegas da área de ensino, dado que, nesta comunidade, como pode ser visto na Tabela~\ref{tab:eventos}, a representação feminina é maior que em toda a comunidade da SBF, com 38\% no SNEF (maior evento nacional na área de ensino) e
41\% no EPEF (sobre pesquisa nessa área), sendo ambos bienais, e com 35\% no ENFNNE-Ensino, realizado entre 2010 e 2015.

No caso do ENFPC, 18\% de participantes de gênero feminino encontra-se um pouco abaixo do percentual de sócias, mostrado na Fig.~\ref{fig:socias}, enquanto nos  demais encontros temáticos e no regional (EFNNE), a proporção feminina está na faixa de 23 a 25 \%, consistente com o percentual de sócias. 

No caso das escolas, essa proporção apresenta uma variação mais ampla, entre 13 e 36 \%. Cabe notar que o tamanho das escolas é tipicamente menor que o dos encontros, e a comunidade de participantes é constituída majoritariamente por estudantes, em número variável.

\section{Composição por atividade desenvolvida no evento}

  Para os diversos eventos, além da proporção feminina dentre os participantes (de todas as formas de participação), consideramos a proporção feminina nos comitês e na lista de palestrantes convidados.
O conjunto de membros de comitês engloba 
todos os tipos de participação em comitês: coordenadores (geral e de programa), comitês de áreas, regionais e outros, sendo que a estrutura de comitês pode mudar de um evento para outro e de uma edição para outra do mesmo evento. 
Palestras convidadas incluem plenárias e 
paralelas de área.  Os demais tipos de apresentações orais e posters não foram contabilizados.

  \begin{figure}[b!]
\centering
\includegraphics[width=0.6\textwidth]{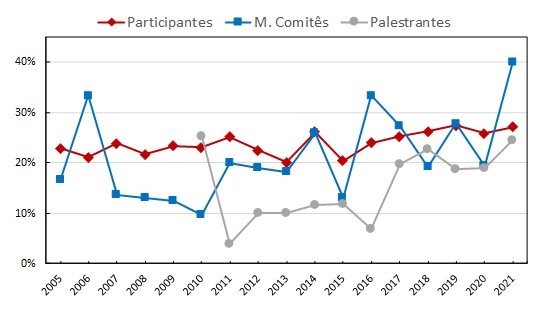}
\caption{Percentual feminino nos diferentes grupos de atuação no ENFMC-EOSBF, indicados na legenda, em cada ano do evento. Falta a informação sobre palestrantes antes de 2010.
}
\label{fig:EOSBF}
\end{figure}

Na Fig.~\ref{fig:EOSBF}, mostramos resultados para o 
ENFMC-EOSBF.
O Encontro Nacional de Física da Matéria Condensada (ENFMC), que passou a englobar outras áreas e a se denominar Encontro de Outono da SBF (EOSBF) desde 2017~\cite{EOSBF}, é um dos mais tradicionais eventos da SBF. 
Atualmente reúne as seguintes áreas:
física atômica e molecular, 
biológica, estatística e computacional, 
da matéria condensada e materiais, 
médica, física na empresa, ótica e fotônica. A física de plasmas também participa do EOSBF, a cada dois anos.
Mostramos nesta figura os percentuais femininos nos diferentes conjuntos (todos os participantes, membros de comitês e palestrantes) do ENFMC-EOSBF.
Neste caso, o percentual feminino de  participantes reflete a representatividade do grupo dentre os sócios da SBF. 
A participação feminina em comitês, excetuando o pico em 2006, tem sido sistematicamente menor que a total no evento, até aprox. 2013, e posteriormente flutua em torno do percentual de participantes gerais. 
Palestrantes de gênero feminino são ainda mais sub-representadas, em percentual de aproximadamente metade daquele de  participantes, até 2016, quando o percentual tende a se aproximar do da população feminina, mas ainda permanecendo menor.

\begin{figure}[h!]
\centering
\includegraphics[width=0.6\textwidth]{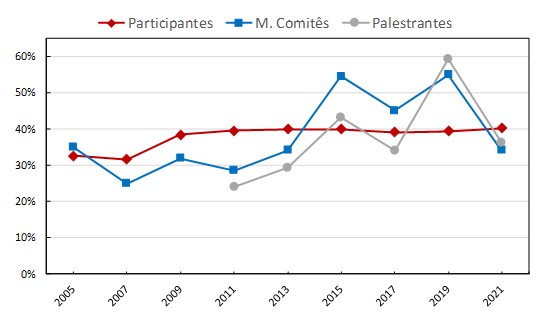}
\caption{Percentual feminino nos diferentes grupos do SNEF, indicados na legenda, em cada ano do evento. Falta a informação sobre palestrantes antes de 2011.}
\label{fig:SNEF}
\end{figure}

Para o SNEF, da área de ensino, a mesma análise anterior é apresentada na Fig.~\ref{fig:SNEF}. Como vimos antes com relação à tabela~\ref{tab:eventos}, 
na área do evento, a participação feminina é maior que em outras áreas, tendo aumentado e se mantido  próxima de 40\% nos últimos anos. 
Entretanto, até 2013, também neste caso houve uma proporção menor de participantes da organização e menor ainda de palestrantes, com relação ao percentual de participantes.  
Já depois de 2013 os números flutuam em torno ou acima dos percentuais de participantes.  

\begin{figure}[h!]
\centering
\includegraphics[width=0.55\textwidth]{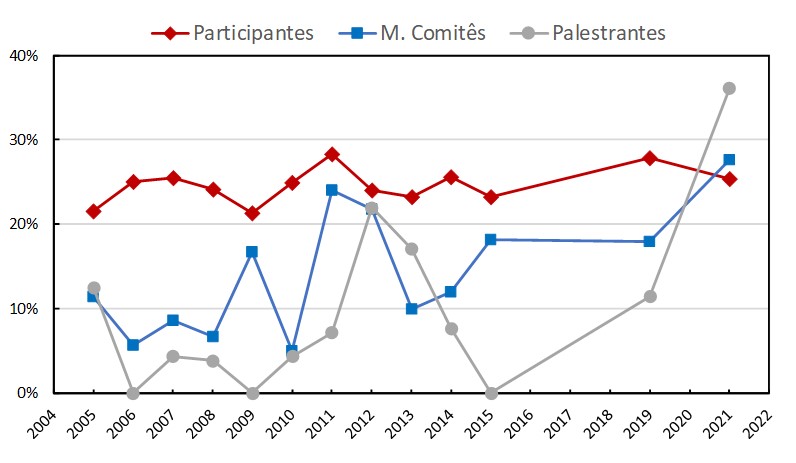}
\caption{Percentual feminino nos diferentes grupos de atuação no EFNNE, indicados na legenda, em cada ano do evento. }
\label{fig:EFNNE}
\end{figure}

O caso do evento regional EFNNE é apresentado na Fig.~\ref{fig:EFNNE}, 
vemos uma participação feminina baixa em comitês e como palestrantes ao longo dos anos, mas houve uma melhora na edição de 2021.

 \begin{figure}[h!]
\centering
\includegraphics[width=0.55\textwidth]{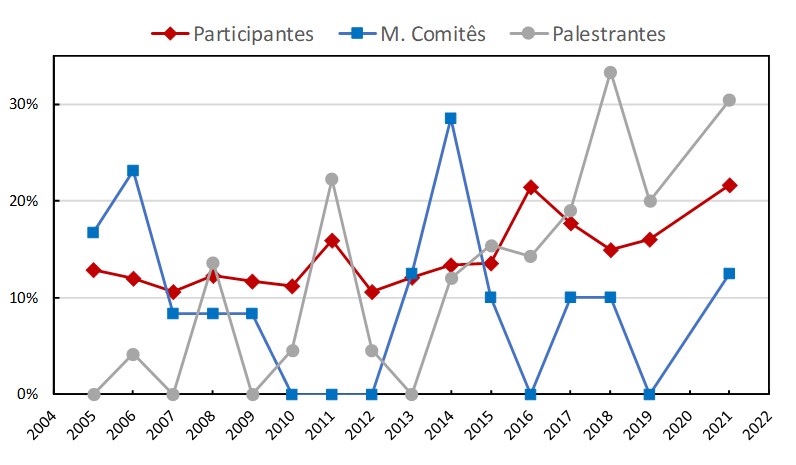}
\caption{Percentual feminino nos diferentes grupos do ENFPC, indicados na legenda, em cada ano do evento. }
\label{fig:ENFPC}
\end{figure}
 \begin{figure}[h!]
\centering
\includegraphics[width=0.55\textwidth]{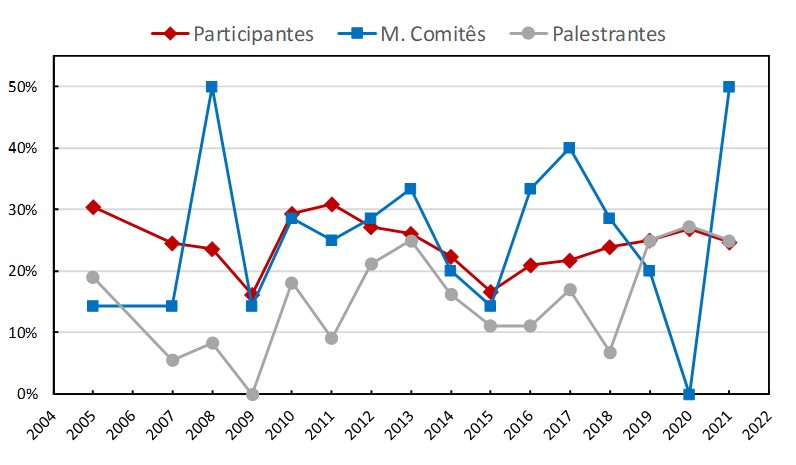}
\caption{Percentual feminino nos diferentes grupos da RTFNB, indicados na legenda, em cada ano do evento. }
\label{fig:RTFNB}
\end{figure}

Os casos do  ENFPC (partículas e campos) e da RTFNB (nuclear) são apresentados na Figs.~\ref{fig:ENFPC} 
e ~\ref{fig:RTFNB}, respectivamente. 
Estes são eventos de menor porte, e portanto, com maiores flutuações. Porém, observamos sistematicamente    uma baixa participação feminina nos comitês do ENFPC e dentre os palestrantes na RTFNB.

\section{ Discussão}

Neste trabalho    quantificamos  a participação feminina em eventos  científicos da SBF, e  observamos um desbalanço especialmente quando se trata de atividades de gestão de eventos e de destaque como palestrantes. A participação feminina em comitês de eventos é menor que a masculina.  Entretanto, nos anos mais recentes, tem havido uma participação que se aproxima da de filiadas à SBF. 
Porém, a proporção de palestrantes do grupo feminino, apesar de ter aumentado nos últimos anos, continua baixa, inclusive em comparação com o percentual na comunidade. 
Estas observações são mais acentuadas fora da área de ensino, onde o grupo feminino está melhor representado.

 Estas constatações não são exclusividade da área de física nem é um caso isolado observado apenas no Brasil. A falta de diversidade de gênero em eventos científicos foi identificada em alguns trabalhos~\cite{Debarre_2018, kibbe_2020, Arora_2020, Platoni_2018, Shishkova_2017} e é um problema por várias razões. Uma delas é que ministrar palestras e principalmente ser palestrante convidada são reconhecimentos acadêmicos e são relevantes na progressão da carreira. Ter menos oportunidade de expor o seu trabalho é portanto um dos mecanismos que leva à redução de mulheres no decorrer da carreira, conhecido como Efeito Tesoura~\cite{Bezerra_2017, Menezes_2017, reportElsevier_2019} e contribui para o teto de cristal. 
Uma outra razão, não menos relevante, é que um leque menor de pontos de vista recebe visibilidade.
 Portanto, a nossa análise é importante para fazer com que a comunidade de física fique atenta ao problema e busque realizar ações tencionando corrigir a baixa participação feminina, e outras disparidades de gênero, em eventos organizados pela SBF.  

Vários estudos identificam uma correlação entre a proporção feminina entre os organizadores dos eventos e a proporção feminina entre os palestrantes convidados~\cite{Debarre_2018, kibbe_2020, Arora_2020, Casadevall_2014}. Em particular, em \cite{Casadevall_2014} mostra-se que a presença de pelo menos uma mulher na equipe de organização tem correlação com uma proporção significativamente maior de oradoras convidadas e menor probabilidade de haver eventos com convidados exclusivamente masculinos. Estes achados apontam para a necessidade de uma composição mais diversa, em todos os aspectos, nas organizações dos eventos científicos.  

A sugestão acima pode levar a pensar que este é um problema que só pode ser resolvido por mulheres, mas este não é o caso. 
\'E crucial que o problema seja reconhecido por toda a comunidade e, que todos trabalhem conjuntamente para torná-la mais inclusiva e diversa. 
Algumas atitudes em nível individual  podem contribuir para mudar o cenário.  Cada interessado e potencial participante do evento pode ficar atento, avaliar o nível de diversidade, e se necessário, contatar os organizadores sugerindo novos nomes para palestrantes e membros de comitês. Esta atitude, além de contribuir diretamente a aumentar a diversidade e inclusão, transmite a mensagem de que este aspecto é considerado relevante pela comunidade. 

 Vale notar que parte dos esforços para aumentar a participação feminina pode ser limitada pelo fato de que as convidadas mulheres  tendem a demorar mais para responder e rejeitam os convites mais seguidamente do que os homens, segundo observado por algumas pessoas da comunidade de física que participam de organização de eventos, mas não  temos estatísticas a respeito. A nossa hipótese para explicar esta observação é que há proporcionalmente poucas mulheres que alcançam os níveis mais elevados na carreira e o decorrente reconhecimento profissional~\cite{reportElsevier_2019}. Assim, essas poucas mulheres são frequentemente mais sobrecarregadas,  enquanto outras pesquisadoras mais jovens ou menos conhecidas não são incentivadas. Uma maneira de contornar este problema é construir uma lista de mulheres a serem convidadas e pedir que as pessoas da comunidade contribuam para montar um banco de dados.  Incluir jovens pesquisadoras com grande potencial e excelentes contribuições  permite  dar espaço a mais mulheres e dividir o encargo com as que não tem mais espaço  na agenda. 

Ações institucionais também podem ser realizadas. Sociedades científicas deveriam elaborar (as que ainda não têm) um guia de diretrizes de igualdade de oportunidades  à exemplo do que existe em algumas sociedades como a Sociedade Brasileira de Física~\cite{SBF_eventos, SBF_etica} e a American Physical Society~\cite{APS_ethics}. 
Uma vez elaborado este guia, contendo sugestões de como organizar eventos mais diversos,  e outras informações,   deve ser constantemente discutido e divulgado para tornar clara a valorização da equidade de oportunidades na ciência.  
Também é crucial que agências de fomento e organizações científicas que financiam eventos explicitem em seus editais de auxílio   a necessidade de diversidade em tais eventos, a exemplo do que faz a  IUPAP (The International Union of Pure and Applied Physics)~\cite{iupap}.

Finalmente, é claro que estas ações para aumentar a diversidade devem ser aplicadas igualmente a outros recortes da comunidade além de sexo e gênero, por exemplo, recortes étnico, geográfico, institucional, etário, etc..

\section{Agradecimentos}
Os dados mostrados neste artigo foram levantados pelos funcionários da SBF, Fernando Braga, Roberto Carvalho e Michele Brizolla, a quem agradecemos.


\begin{thebibliography}{99}



\bibitem{Gutterl_2014}
F. Gutterl, {\it  Diversity in science: Why it is essential for
excellence}. Sci Am 311: 38-41 (2014).

\bibitem{McKinsey}
V. Hunt, D. Layton, S. Prince, {\it Diversity
Matters: McKinsey and Company Report} (2015). 
\url{http://www.mckinsey.com/business-functions/organization/our-insights/why-diversity-matters}. (Accessed on Jul 19, 2022).

\bibitem{Philips_2014}
K. W. Philips, {\it How diversity makes us smarter}, Sci Am 311: 42-47 (2014).

\bibitem{wooley15} 
A. W. Woolley, I. Aggarwal, T. W. Malone, {\it Collective
intelligence and group performance}, Curr. Dir. Psychol. Sci. 24, 420 (2015).


\bibitem{PRPER}  
C. Anteneodo, C. Brito, A. Alves-Brito, S. S. Alexandre, B. Nattrodt D’Avila, D. Peres Menezes, 
{\it Brazilian physicists community diversity, equity, and inclusion: A first diagnostic}, 
Physical Review Physics Education Research 16, 010136 (2020).

\bibitem{eventosSBF}
\url{http://www.sbfisica.org.br/v1/home/index.php/pt/eventos/eventos-realizados}, acessada em 22/2/22. 

\bibitem{EOSBF}
\url{http://www1.fisica.org.br/~enfmc/xli/index.php/pt/index.html}


\bibitem{Debarre_2018}  
F. Débarre, N. O. Rode, L. V. Ugelvig, {\it Gender equity at scientific events}, Evolution Letters 2-3: 148–158 (2018).

\bibitem{kibbe_2020}  
 M. R. Kibbe, M. R. Kapadia {\it Underrepresentation of Women at Academic Medical Conferences—“Manels” Must Stop}, JAMA Netw Open. 2020; 3(9):e2018676.

 
\bibitem{Arora_2020}  
A. Arora, Y. Kaur, F. Dossa, R. Nisenbaum, D. Little, N. N. Baxter, {\it Proportion of Female Speakers at Academic Medical Conferences Across Multiple Specialties and Regions}, JAMA Network Open. 2020;3(9):e2018127.

\bibitem{Platoni_2018}
K. Platoni, S. Triantopoulou, M. Dilvoi, E. Koutsouveli, A. Ploussia, V. Tsapaki, {\it Participation of women medical Physicists in European scientific events: The European experience}, Physica Medica
Volume 46, Pages 104-108 (2018).

\bibitem{Shishkova_2017}
E. Shishkova, N.  W. Kwiecien, A. S. Hebert, M. S. Westphall, J. E. Prenni,  J. J. Coon, 
{\it Gender Diversity in a STEM Subfield – Analysis of a Large Scientific Society and Its Annual Conferences} J. Am. Soc. Mass Spectrom., 28, 12, 2523–2531 (2017).


\bibitem{Bezerra_2017}
 G. Bezerra, M. Barbosa, {\it Mulheres na física no Brasil:
Contribuição de alta relevância, mas, por vezes, ainda
invisível}, in SBF: 50 Anos, pp. 130–133 (2017), 
\url{http://www.sbfisica.org.br/arquivos/SBF-50-anos.pdf}

\bibitem{Menezes_2017}
D. P. Menezes, C. Brito, C. Anteneodo, {\it Women in
physics: Scissors effect from the Brazilian Olympiad of
physics to professional life}, arXiv:1901.05536 and Sci. Am.
Brasil, 76 (2017).
 

\bibitem{reportElsevier_2019}
Gender in the Global Research Landscape Report,
Elsevier, London, 2019, 
\url{https://www.elsevier.com/research-intelligence/resource-library/gender-report}


\bibitem{Casadevall_2014}  
A Casadevall  J. Handelsman, {\it The presence of female conveners correlates with a higher proportion of female speakers at scientific symposia}, mBio 5(1):e00846-13 (2014).

\bibitem{SBF_eventos}
\url{http://www.sbfisica.org.br/v1/home/images/eventos/recomendacoes-para-eventos-e-codigo-de-conduta.pdf}

\bibitem{SBF_etica}
\url{https://www.sbfisica.org.br/v1/home/index.php/pt/eventos/codigo-de-conduta}

\bibitem{APS_ethics}
\url{https://www.aps.org/policy/statements/guidlinesethics.cfm}

\bibitem{iupap}
\url{https://iupap.org/conferences/conference-policies/}


\end{thebibliography}

\end{document}